\documentclass[fleqn,usenatbib]{mnras}
\usepackage[utf8]{inputenc}
\usepackage{amsmath}
\usepackage{amssymb}
\usepackage{tabularx}
\usepackage{breqn}
\usepackage{appendix}
\usepackage{xcolor}

\definecolor{seagreen}{rgb}{0.18, 0.55, 0.34}

\title{Dispersion Measures of Fast Radio Bursts through the Epoch of Reionization}
\author[J. Ziegler et al.]{Joshua J. Ziegler,$^{1}$ \thanks{jjziegler@utexas.edu}
Paul R. Shapiro,$^{2}$ 
Taha Dawoodbhoy,$^{3}$ 
Paz Beniamini,$^{4,5,6}$ 
Pawan Kumar,$^{2}$ 
\newauthor
Katherine Freese,$^{1,7,8}$
Pierre Ocvirk,$^{9}$
Dominique Aubert,$^{9}$
Joseph S. W. Lewis,$^{9}$ 
Romain Teyssier,$^{10}$
\newauthor
Hyunbae Park,$^{11}$ 
Kyungjin Ahn,$^{12}$
Jenny G. Sorce,$^{13,14,15}$ 
Ilian T. Iliev,$^{16}$ 
Gustavo Yepes,$^{17,18}$
\newauthor
Stefan Gottl\"ober$^{15}$ 
\\
$^{1}$Department of Physics, The University of Texas at Austin, Austin, Texas 78712, USA\\
$^{2}$Department of Astronomy, The University of Texas at Austin, Austin, Texas 78712, USA\\
$^{3}$Department of Physics, California Polytechnic State University, San Luis Obispo, California 93407, USA\\
$^{4}$ Astrophysics Research Center of the Open University (ARCO), The Open University of Israel, P.O Box 808, Ra’anana 4353701, \\Israel\\
$^{5}$Department of Natural Sciences, The Open University of Israel, P.O Box 808, Ra’anana 4353701, Israel\\
$^{6}$Department of Physics, The George Washington University, 725 21st Street NW, Washington, DC 20052, USA\\
$^{7}$The Oskar Klein Centre, Department of Physics, Stockholm University, AlbaNova, SE-10691 Stockholm, Sweden\\
$^{8}$Nordic Institute for Theoretical Physics (NORDITA), 106 91 Stockholm, Sweden\\
$^{9}$CNRS, Observatoire astronomique de Strassbourg, Universit\'e de Strasbourg, UMR 7550, F-67000 Strasbourg, France\\
$^{10}$Department of Astrophysical Sciences, Princeton University, 4 Ivy Lane, 08540, Princeton, NJ, USA\\
$^{11}$Center for Computational Sciences, University of Tsukuba, 1-1-1 Tennodai, Tsukuba, Ibaraki 305-8577, Japan
 \\
$^{12}$Chosun  University, 375 Seosuk-dong, Dong-gu, Gwangjiu 501-759, Korea\\
$^{13}$Univ. Lille, CNRS, Centrale Lille, UMR 9189 CRIStAL, F-59000 Lille, France\\
$^{14}$Universit\'e Paris-Saclay, CNRS, Institut d'Astrophysique Spatiale, 91405, Orsay, France\\
$^{15}$Leibniz-Institut fur Astrophysik Potsdam (AIM), An der Sternwarte 16, D-14482 Potsdam, Germany\\
$^{16}$Astronomy Center, Department of Physics \& Astronomy, University of Sussex, Pevensey II Building, Falmer, Brighton BNI 9QH, \\UK\\
$^{17}$Departmento de F\'isica Te\'orica M-8, Universidad Aut\'onoma de Madrid, Cantoblanco, E-28049 Madrid, Spain\\
$^{18}$Centro de Investigaci\'on Avanzada en F\'isica Fundamental (CIAFF), Facultad de Ciencias, Universidad Aut\'onoma de Madrid, \\Cantoblanco, E-28049 Madrid, Spain}

\date{Accepted XXX. Received YYY; in original form ZZZ}

\pubyear{2024}

\begin{document}

\label{firstpage}
\pagerange{\pageref{firstpage}--\pageref{lastpage}}

\maketitle

\begin{abstract}
Dispersion measures (DM) of fast radio bursts (FRBs) probe the density of electrons in the intergalactic medium (IGM) along their lines-of-sight,
including the average density versus distance to the source and its variations in direction. While previous study focused on low-redshift, FRBs are potentially detectable out to high redshift, where their DMs can, in principle, probe the epoch of reionization (EOR) and its patchiness.  We present the first predictions from large-scale, radiation-hydrodynamical simulation of fully-coupled galaxy formation and reionization, using Cosmic Dawn (``CoDa")~II to model the density and ionization fields of the universe down to redshifts through the end of the EOR at $z_{re}\approx6.1$. Combining this with an N-body simulation CoDa~II--Dark Matter of the fully-ionized epoch from the EOR to the present, we calculate the mean and standard deviation of FRB DMs as functions of their source redshift. The mean and standard deviation of DM increase with redshift, reaching a plateau by $z(x_{HII}\lesssim0.25)\gtrsim8$, i.e. well above $z_{re}$. The mean-DM asymptote $\mathcal{DM}_{max} \approx 5900~\mathrm{pc\, cm^{-3}}$ reflects the end of the EOR and its duration. The standard deviation there is $\sigma_{DM, max}\approx497 ~\mathrm{pc\, cm^{-3}}$, reflecting inhomogeneities of both patchy reionization and density. 
Inhomogeneities in ionization during the EOR contribute $\mathcal{O}(1$ per cent) of this value of $\sigma_{DM,max}$ from FRBs at redshifts $z\gtrsim 8$.
Current estimates of FRB rates suggest this may be detectable within a few years of observation. 
\end{abstract}


\maketitle

\section{Introduction}

Photons emitted during a fast radio burst (FRB) with different wavelengths undergo different amounts of plasma dispersion, causing them to be observed at different times. By measuring the spread of these arrival times, it is possible to determine the integrated electron density along the line of sight (LOS) from the FRB source to observation on Earth\citep{Petroff_2019, Cordes_2019}.
Due to their high luminosity, FRBs are detectable out to great distance and correspondingly large look-back time, with some 50 FRBs localized to source galaxies at redshifts up to $z=1.017$ \citep{Lorimer_2024, gordon2023fast} and some hundreds to thousands of unlocalized FRBs including several believed to be produced at higher redshifts\citep{Lorimer_2024}. Furthermore, while the sources of FRBs are not fully understood, most models identify black holes or neutron stars (especially magnetars) as likely candidates \citep{Rees1977, Vieyro_2017, Gupta_2018, Popov&Postnov13, Katz16, Kumar+17, KumarBosnjak2020, Margalit_2018, Wadiasingh2019,BK2023}. 
Because both black holes and neutron stars form primarily through the supernova of massive stars, FRBs may be present wherever stars form. If this is the case, measurements of the dispersion measures of FRBs can probe the density of ionized gas throughout most of the universe's history, from the formation of the first stars and the epoch of reionization (EoR)  to the present. In addition, since the stars believed responsible for most of the emission of the UV radiation that caused reionization are also the stars that undergo the supernovae that produce neutron stars and black holes \citep{Wise_2019}, it is a natural expectation that FRBs were emitted during the EoR.

For distant FRBs, the intergalactic medium (IGM)\footnote{When discussing dispersion measures, it is often practical to consider contributions from the Milky Way, source galaxies, and everything in between. This latter category is what we identify as IGM, even though it also includes any haloes intervening between the source and Milky Way galaxies.} is expected to be the dominant contributor to the dispersion measure \citep{Zhu_2021}, and as such, studying the dispersion measures of these distant FRBs could allow us to probe the ionization history of the IGM. In particular, statistical variations in the dispersion measure could tell us something about the rise of structure in the universe and the resulting inhomogeneous nature of reionization \citep{Beniamini2021}. The simplest case would be if sufficient numbers of FRBs could be localized to a host galaxy so that their redshifts were known. However, even without FRB redshifts, identifiable properties in the distribution of dispersion measures could also be used. In either case, observations of FRB dispersion measures could be combined with modeling of the theoretical dependence of FRB dispersion measures on redshift to better understand and constrain the patchiness of reionization and the growth of structure more generally. 

In what follows, we compute the dispersion measure through the universe along different LOS's, based upon cosmological simulations, to determine the mean and variance of this quantity as functions of redshift. As we are most interested in improving upon previous work by taking proper account of the contribution from the EoR, including its reionization patchiness, for the first time, we take advantage of a recent fully-coupled, large-scale, high-resolution, radiation-hydrodynamical simulation of galaxy formation and reionization during the EoR by the Cosmic Dawn (``CoDa'') Project, the CoDa~II simulation \citep{Ocvirk_2020}, down to redshift 5.8, past the end of the EoR
\footnote{The specific definition used for the end of the EoR is not universal, but is usually based upon a certain value of the globally-averaged neutral fraction of intergalactic hydrogen.  In CoDa II, the EoR ends at $z=6.1$, where the neutral fraction ends its sharp, exponentially-steep drop from $0.99$ at $z=6.2$ to $10^{-4.6}$ at $z=6.1$. \citep{Ocvirk_2020}},
for our analysis. By this redshift, the universe was almost fully ionized and, to a good approximation, may be assumed to be fully ionized, thereafter, with the matter density field and the ionized gas density field tracing each other on large scales. To extend the LOS integrals of dispersion measure from $z=5.8$ to $z=0$, therefore, we use a dark-matter-only companion simulation to CoDa~II, an N-body simulation called CoDa~II--Dark Matter, which was run with the same initial conditions as CoDa~II\citep{Ocvirk_2020}.

Previous work applied earlier cosmological simulations of varying degrees of physical realism to compute the FRB dispersion measures and their properties. Most of these previous analyses calculate dispersion measures from simulation results to redshifts below $z \simeq 6$, preventing them from probing the EoR. 
\citet{McQuinn_2013} used the results of a simulation by \citet{Faucher_Giguere_2011} to model the mean and standard deviation of the dispersion measure out to redshift 1.4, and compared this to predictions from dark-matter-only halo density functions and simple top-hat models of structure formation. \citet{Dolag2015} used the Magneticum Pathfinder simulations to calculate the dispersion measures out to redshift 2. \citet{Jaroszynski_2019} used the results of the Illustris-3 simulation \citep{Nelson_2015} to calculate dispersion measures out to redshift 5. \citet{Pol_2019} used the results of the MICE simulation \citep{Fosalba_2008} to calculate dispersion measures out to redshift 1.4. \citet{Batten2021} used the EAGLE simulation \citep{Schaye_2015} to calculate dispersion measures out to redshift 3. \citep{Walker2023} used the IllustrisTNG simulation \citep{Nelson2019} to calculate dispersion measures out to redshift 5. Each of these simulation efforts involve different gas and dark matter mass resolutions, as well as different simulation volume sizes, which may contribute to slightly different results.
In contrast to the studies detailed above, \citet{Zhang_2021} do calculate the dispersion measure to a redshift of $z=9$ using IllustrisTNG-300 data, and consequently are able to explore the EoR. Our current analysis differs from this previous work in the treatment of radiation in the two underlying simulations. Because the CoDa~II simulation models UV radiation self-consistently, we can study the effects of the spatial inhomogeneities (patchiness) on the FRB dispersion measure.  We also consider a larger range of redshift (out to $z\gtrsim12$).

In this paper, we compute the mean and standard deviation of the dispersion measure to FRBs as a function of the source redshift of those FRBs. In section~\ref{theory}, we discuss the dispersion measure of cosmologically distant FRBs, and in section~\ref{methods}, we describe how we can use the results of the CoDa~II simulations to compute these dispersion measures. We present our results in section~\ref{results}, particularly focusing on the mean, standard deviation, and ratio of mean to standard deviation of the dispersion measures we calculate. The inset of Fig.~\ref{fig:DMstds} shows one of our primary results, namely that the patchiness/inhomogeneity of ionization during the EoR contributes an effect that is distinguishable at the $\mathcal{O}(1$ per cent) level among FRBs from sources at redshifts $z\gtrsim 8$. Finally, in section~\ref{toys}, we discuss how our results might change if the ionization history assumed in the CoDa~II simulations were to be altered. We conclude in section~\ref{conc}.

\section{Arrival time delay of FRB signals at cosmological distances}
\label{theory}
The presence of free electrons of number density $n_{e}$ along the line of sight (LOS) to a distant FRB introduces a frequency-dependent time delay, $t_{d}(\nu)$, in the arrival of the FRB signal observed at frequency $\nu$, relative to its arrival time in vacuum, due to plasma dispersion. If the source is located a distance $D$ from the observer, in a static medium and in the absence of cosmological redshift, the rate of change of this time delay with respect to observed frequency, is given by

\begin{equation}
\frac{d t_{d}}{d \nu}=-\frac{e^{2}}{\pi m_{e} c} \frac{\mathcal{DM}}{\nu^{3}} 
\label{dtdnu}
\end{equation}
where the dispersion measure $\mathcal{DM}$ is defined by \cite{Petroff_2019}

\begin{equation}
\mathcal{DM} \equiv \int_{0}^{D} n_{e} d l 
\label{DMdef}
\end{equation}
where $\mathrm{dl}$ is the differential distance along the LOS.

For a source at cosmological distance, however, the derivative of time delay with respect to observed frequency is modified to account for the effects of cosmic expansion. Specifically, we account for changes in photon frequency, path length, and electron density. We also account for cosmological time dilation, where a time interval at the FRB source will be stretched due to cosmic expansion by the time it reaches the observer.

Consider an FRB source at redshift $z_s$, which is observed at redshift $z=0$ and direction on the sky $\mathbf{\hat{n}}$ to have a frequency $\nu$. As a result of cosmic expansion, the frequency $\nu^\prime$ emitted at $z=z_s$ is relate to this observed frequency $\nu$ as
\begin{equation}
\nu^{\prime}=\nu(1+z). 
\label{redshift}
\end{equation}
In addition, over each interval of cosmic time $dt$, the photon travels a proper distance $dl$ given by 
\begin{equation}
d l=c d t=c\left|\frac{d t}{d z}\right| d z. 
\label{time position}
\end{equation}
The quantity $|\frac{dt}{dz}|$, relating cosmic time and redshift intervals, is given by
\begin{equation}
\left|\frac{d t}{d z}\right|=\frac{1}{(1+z) H(z)}, 
\label{dtdz}
\end{equation}
where the Hubble parameter $H(z)$ is (for redshifts significantly below matter/radiation equality)
\begin{equation}
H^{2}=H_{0}^{2}\left[\Omega_{m,0}(1+z)^{3}+\Omega_{\Lambda}\right] .
\label{Hubbleparam}
\end{equation}
Here, $H_0$ is the local Hubble constant, and $\Omega_{m,0}$ and $\Omega_\Lambda$ are the matter and dark energy density parameters $\rho/\rho_c$ evaluated at redshift $z=0$.

\begin{figure}
    \centering
    \includegraphics[width=\columnwidth]{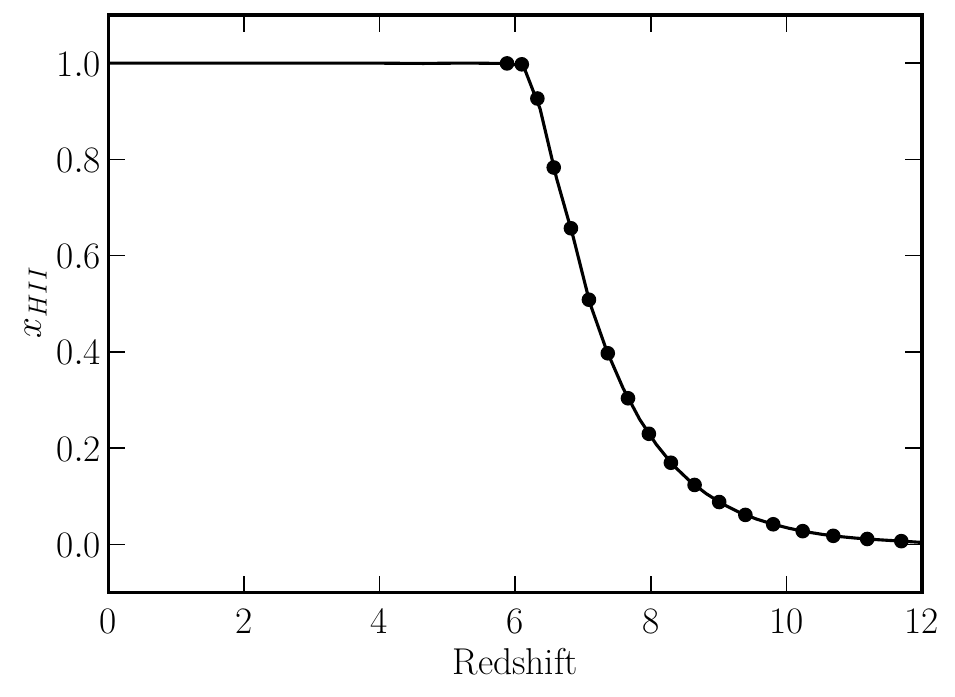}
    \caption{Globally-averaged H ionized fraction, 
    $\langle x_{HII}\rangle $, vs. redshift $z$, 
    from the Cosmic Dawn II simulation. 
    Black points show the hydrogen ionization fraction of snapshots at different redshifts from the CoDa~II simulation, averaged over the simulation volume. The curve connecting them represents a cubic spline fit to these points. In this model, reionization is largely completed around redshift $z\approx6$, and the universe is half-ionized around redshift $z\approx 7$.
    }
    \label{fig:ionization}
\end{figure}

The free electron number density $n_{e}(z, \hat{n})$ at redshift $z$ along the LOS in direction $\hat{n}$ is related to the mean number density of $\mathrm{H}$ atoms at present, $n_{H, 0}$, the number of free electrons per $\mathrm{H}$ atom, $x_{e}$, and the fractional overdensity, $\delta_{H}$, of $\mathrm{H}$ atoms at that location along the LOS, according to
\begin{equation}
n_{e}(z, \mathbf{\hat{n}})=x_{e}(z, \mathbf{\hat{n}}) n_{H, 0}(1+z)^{3}\left(1+\delta_{H}(z, \mathbf{\hat{n}})\right). 
\label{electrondensity}
\end{equation}
The mean number density of hydrogen atoms at redshift $z=0$ is defined as 
\begin{equation}
n_{H, 0}=\left(\frac{3 {H_{0}}^{2} \Omega_{b,0}}{8 \pi G m_{p}}\right)X,
\label{Hatomdensity}
\end{equation}
where $\Omega_{b,0}$ is the present density parameter of baryons and $X$ is the mass fraction of the baryons that are hydrogen.
The number of free electrons will vary over redshift as a function of the fraction of hydrogen atoms that have ionized and of the fraction of helium atoms that have ionized singly or doubly:
\begin{equation}
    x_e = \frac{1}{X}\left[Xx_{HII}(\mathbf{\hat{n}},z) + \frac{1}{4}Yy_{HeII}(\mathbf{\hat{n}},z) + \frac{1}{2}Yy_{HeIII}(\mathbf{\hat{n}},z)\right].
\end{equation}
Here, $X=0.76$ and $Y=0.24$ are the mass fractions of hydrogen and helium, respectively, and the three fractions 
$x_{}, y_{HeII}$, and $y_{HeIII}$ are the ionization fraction of hydrogen, singly ionized helium, and doubly ionized helium, respectively. We assume that the second ionization of helium occurs instantaneously at redshift $z=3$. Therefore, prior to $z=3$, $y_{HeIII}=0$ and it is expected that $x_{HII} \simeq y_{HeII}$.
Conversely for $z \leq 3$, $x_{HII}=y_{HeIII}=1$. Throughout this paper, we present ionization fractions as the hydrogen ionization fraction $x_{HII}$ (which CoDa papers refer to as the mass-weighted ionized fraction $x_m$), and use these relations between the hydrogen and helium ionization fractions to compute the ionization fraction of helium. 
Using this convention, we can relate the hydrogen ionization fraction to the electron fraction per hydrogen atom $x_e$ as 
\begin{equation}
    x_e = \begin{cases}
    1.16 x_{HII} & z\leq 3\\
    1.08 x_{HII} & z>3
    \end{cases}.
\end{equation}
The history of the globally-averaged H ionization fraction in the CoDa II simulation is plotted in Fig.~\ref{fig:ionization}.

Finally, due to cosmological time dilation, if a time delay of $d t_{d,z}$ were measured by an observer at redshift $z$, even without any free electrons at redshifts less than $z$, an observer at redshift $z=0$ would measure a time delay 
\begin{equation}
d t_{d, r e c}=(1+z) d t_{d, z}.
\label{dtrecdt}
\end{equation}

The differential arrival delay per observed frequency interval $d \nu$ at observed frequency $\nu$ for a cosmologically distant source that bursts at redshift $z_s$, is then given by multiplying $d t_{d}$ in equation (\ref{dtdnu}) by $(1+z)$, to make it $d t_{d, r e c}$, and substituting equations (\ref{time position}) - (\ref{Hatomdensity}) into the right-hand-side of equation (\ref{dtdnu}). The cumulative time delay $\Delta t_{d, \nu}$ observed at present at a given frequency $\nu$ from a source at redshift $z_{s}$ in direction $\hat{n}$, relative to the arrival time in the absence of free electrons, is then given by integrating over the contributions from different redshifts along the LOS between the source at $z_s$ and the origin at $z=0$.  This can still be expressed in terms of the familiar expression for (noncosmological) time delay as proportional to the dispersion measure, namely,
\begin{equation}
 \Delta t_{d, r e c}\left(z_{s},\hat{n}\right)=\frac{e^{2}}{2 \pi m_{e} c} \frac{\mathcal{D} \mathcal{M}}{\nu^{2}} 
 \label{deltat}
\end{equation}
as long as we also redefine the dispersion measure $\mathcal{D M}$ by the following integral over redshift, for a cosmological source which emits the FRB at redshift $z_{s}$ :

\begin{multline}
   \mathcal{D M}\left(z_{s}, \hat{n}\right) \equiv c n_{H, 0}\\ \times \int_{0}^{z_{s}} d z \,\, x_{e}(z, \hat{n})\left[1+\delta_{H}(z, \hat{n})\right]\frac{(1+z)}{H(z)}  
   \label{cosmic dispersion measure}
\end{multline}
Henceforth in what follows, we shall refer to the cosmological quantity defined in equation (\ref{cosmic dispersion measure}) when we use the term `dispersion measure,' so there is no confusion with the noncosmological quantity for a static, nonexpanding medium.

As a comparison for our later results, we can estimate the mean dispersion measure we would expect as a function of redshift, if the universe were fully ionized at all redshifts. In doing so, we use the values of cosmological parameters that we use throughout the paper, based on the choices made in the CoDa~II simulations. In particular, the matter density parameter at the present day $\Omega_{m,0} = 0.307$, the dark energy density parameter $\Omega_\Lambda = 0.693$, the baryonic matter density parameter at the present day $\Omega_{b,0} = 0.048$, and the Hubble constant at the present day $H_0 = 67.7 ~\mathrm{km\,s^{-1}\,Mpc^{-1}}$. With these values, for all redshifts where baryons can be approximated as fully ionized (i.e. at lower redshifts than the end of the EoR), equation~\eqref{cosmic dispersion measure} could be rewritten as
\begin{equation}
 \mathcal{DM} = 1741 ~\mathrm{pc\,cm^{-3}} \int_{0}^{z_s}  \frac{1+z}{\sqrt{(1+z)^3 + 2.26}} dz.   
\end{equation}

\section{Methodology}
\label{methods}
To evaluate the dispersion measure in equation (\ref{cosmic dispersion measure}) for a given line of sight to an FRB, we must know the time- and spatially-varying values of the 3-D ionized fraction and overdensity fields, $x_e(\bf{r},t)$ and 
$\delta_H(\bf{r},t)$. For this, we use the detailed results of the large-scale radiation-hydrodynamical simulation of galaxy formation and reionization CoDa~II \citep{Ocvirk_2020} through the end of reionization, down to $z=5.8$.   For the redshift range from $z=5.8$ to $z=0$, we use the collisionless, N-body simulation CoDa~II--Dark Matter, which was run from the same initial conditions as CoDa~II (i.e. the same linear-perturbation, plane-wave fluctuation density and velocity modes)\citep{Ocvirk_2020}. Some details of the CoDa~II and CoDa~II--Dark Matter simulations can be found in table~\ref{tab:CoDaGadgetParams}. In CoDa~II, the atomic gas density and ionized fraction fields were computed by modelling fully-coupled galaxy formation and reionization self-consistently, solving the gas dynamical conservation equations, along with explicit treatment of radiative transfer of H-ionizing UV starlight and nonequilibrium ionization rate equations.  In CoDa~II--Dark Matter, the dynamics of collisionless dark matter was treated by a standard cosmological N-body method. 
Nevertheless, we can use the results from the CoDa~II--Dark Matter simulation to model the electron density fields over redshifts $z<5.8$. As $z=5.8$ is later than the end of reionization, as computed by the CoDa~II simulation, we can assume that the hydrogen gas is fully ionized at all later redshifts, and the baryonic and electron density fields closely follow the total matter density field of the N-body simulation.

The CoDa~II simulation was performed in a comoving cubic box of size 94.4 cMpc, utilizing $4096^3$ N-body particles of dark matter and a fixed Eulerian grid of $4096^3$ cells for both the baryonic gas and radiation fields. However, to reduce storage requirements, the raw simulation output was smoothed to a coarser grid of $2048^3$ cells. The CoDa~II--Dark Matter simulation, meanwhile, is performed in the same 94.4 cMpc box, but utilized only $2048^3$ particles, and the particle data was used to create a grid-based density field. As a result, the data we used from either simulation's output was in the form of a $2048^3$ grid-based field.

\begin{table*}
    \centering
    \begin{tabular}{|c|c|c|}
        \hline
         Simulation & CoDa~II & CoDa~II--Dark Matter\\
         &  & (CoDa~II--DM2048)\\
         \hline
         Redshift Range & 5.8--150 & 0--150\\
         Evolution Type & baryonic rad-hydro and dark matter gravity & dark-matter-only gravity\\
         Box Length & $94.4$ cMpc & $94.4$ cMpc\\
         Number of Cells (Simulation) & $4096^3$ & $-NA-$\\
         Number of Cells (Analysis) & $2048^3$ & $2048^3$\\
         Number of (Dark Matter) Particles & $4096^3$ & $2048^3$\\
         \hline
    \end{tabular}
    \caption{Comparison of CoDa~II and CoDa~II--Dark Matter Simulation and Analysis Parameters. Throughout the paper, we use two simulations from the CoDa collaboration \protect\citep{Ocvirk_2020}. The simulation labelled as `CoDa~II--Dark Matter' is a dark-matter-only N-body simulation performed using the \textsc{Gadget-2} code. Meanwhile, the simulation labelled as `CoDa~II' includes baryons as well as dark matter, and simulates hydrodynamics, radiative transfer, and nonequilibrium ionization using the \textsc{RAMSES-CUDATON} code. Both of these simulations use the same initial conditions, which were selected to produce an arrangement of galaxies like the Local Group at redshift $z=0$. Some of the numerical parameters used in these two simulations are presented here. Of particular note, while the CoDa~II simulation is performed at a higher resolution ($4096^3$), it was smoothed for analysis to a coarsened resolution consistent with that of the CoDa~II--Dark Matter simulation ($2048^3$). In addition, the CoDa~II simulation was unable to simulate to redshifts below $z=5.8$, while the CoDa~II--Dark Matter simulation was able to continue to redshift $z=0$.} 
    \label{tab:CoDaGadgetParams}
\end{table*}

In order to calculate statistical properties of the dispersion measure from FRBs, we construct multiple lines of sight from the CoDa~II and CoDa~II--Dark Matter simulation results. Specifically, we construct 2 million lines of sight, but the results we describe in the next sections are insensitive to the precise number of lines constructed. We construct each of these lines by following the path a photon would take through density and hydrogen ionization fraction fields generated using the CoDa simulations. Along this path, a photon can travel at most the box length, 94.4 Mpc, before it would sample a patch of the simulation volume that it has sampled previously. Without correction, this would introduce an unphysical correlation between regions of space that are separated by almost 100 Mpc. To minimize these correlations, each time light has traveled 94.4 Mpc, we randomize its direction so that it samples a new set of density and hydrogen ionization fractions. This establishes a natural segmentation of the light travel path into box-length pieces, and we can associate the ends of these segments with redshifts $z_{i,box}$ through equation~\eqref{time position}.  

The CoDa simulations output density and ionization fields across the entire simulation volume at specific times (and corresponding redshifts). This introduces two challenges to directly using the CoDa outputs as the basis for the box-length segments of a photon path. First, the universe's evolution is continuous, while relying on the CoDa snapshots allows only discrete steps in the evolution. This effect increases with redshift, but even at the highest redshifts we consider, it contributes an error of less than 1 per cent. Furthermore, any error like this would be systematic to all of our results, so while it might slightly impact the overall magnitude of the dispersion measures we calculate, the impact on any comparisons will be even smaller. 
As a result, in our calculations, we ignore any effects of the discrete sampling of evolution. With that in mind, we can partially account for the cosmological expansion of the gas by describing the atomic H density at a redshift $z$ within one of the box-length segments as
\begin{equation}
   n_H=n_{H,0}(1+z)^3(1+\delta_H), 
   \label{H density}
\end{equation}
where $\delta_H$ is the overdensity from the nonevolving snapshot data.

\begin{figure}
    \centering
    \includegraphics[width=\columnwidth]{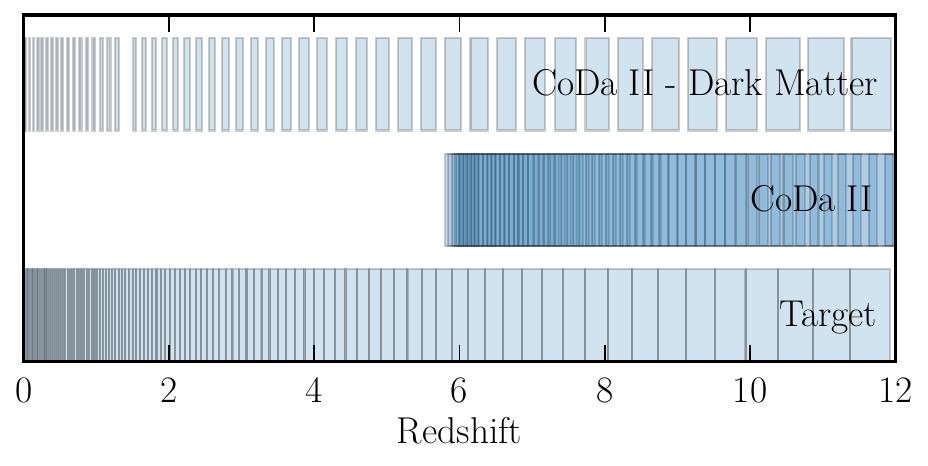}
    \caption{Diagram comparing the time between Cosmic Dawn (CoDa)~II simulation snapshots and those in the CoDa~II--Dark Matter simulation. From the CoDa~II simulation, we averaged over multiple simulated snapshots in the time it takes light to cross any one simulation volume, while from the CoDa~II--Dark Matter simulation, we interpolated between two consecutive simulation snapshots to construct a synthetic snapshot.  Therefore, we approach the process of creating synthetic snapshots based off of these two simulations in slightly different manners, described in detail in the main text and shown diagrammatically in the sequences labelled `CoDa~II' and `CoDa~II--Dark Matter'. Ultimately, our goal in both cases is to generate a series of synthetic snapshots based off the simulation results but spaced in time in such a way that the time between snapshots is exactly the time it takes light to travel across the simulation volume, as depicted in the sequence labelled `Target'. 
    }
    \label{fig:stitching}
\end{figure}

The second challenge introduced by using the outputs from the CoDa simulations is that those outputs are not aligned with the list of redshifts $z_{i,box}$ that define the box-length segments. In particular, the CoDa~II snapshots are recorded more frequently than the segments ($\Delta z_{i,box}>\Delta z_{i,CoDaII}$), while the CoDa~II--Dark Matter snapshots are recorded less frequently ($\Delta z_{i,box}<\Delta z_{i,CoDaIIDM}$). Visually, this comparison can be represented in figure~\ref{fig:stitching}. We treat each of these cases slightly differently, but with the common goal of generating for each box-length segment (between successive redshifts $z_{i,box}$), a distribution of contributions to the dispersion measure. 
More details on this process can be found in appendix~\ref{moremethods}.

\section{Dispersion Measures from the CoDa~II and CoDa~II--Dark Matter Simulations}
\label{results}

\begin{figure}
    \centering
    \includegraphics[width=\columnwidth]{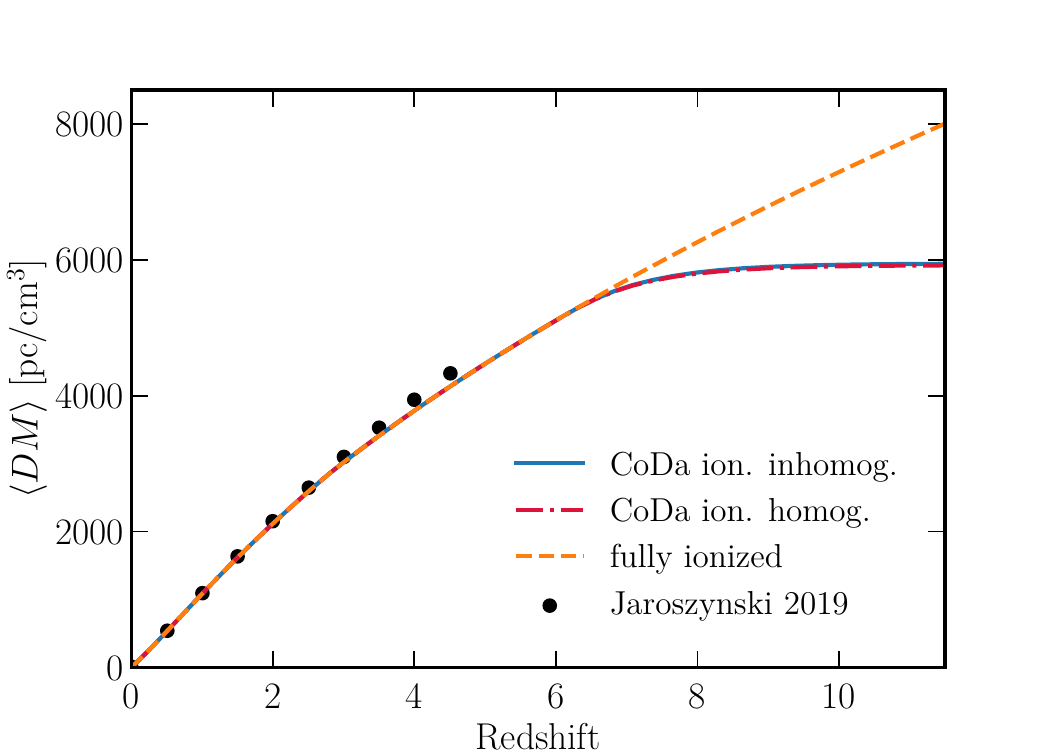}
    \caption{Mean of dispersion measures modeled from CoDa~II/CoDa~II--Dark Matter simulations. The solid blue curve shows the mean of the dispersion measures constructed from the CoDa~II/CoDa~II--Dark Matter simulations of FRBs whose source is located at a range of redshifts $0<z<12$. The orange dashed curve is the mean under the assumption that the universe is fully ionized over that entire redshift range, specifically using the CoDa~II density fields at redshifts $z>5.8$ and CoDa~II--Dark Matter at lower redshifts. The red dash-dotted curve shows the mean if we spatially average the ionization  field from the CoDa~II simulation, while retaining the CoDa~II density field, allowing us to estimate the effects of the globally-averaged rise of the ionized fraction as the universe reionized, treated as a non-patchy (i.e. uniform), evolving, partial ionization.
    The means in these three cases are indistinguishable from $z=0$ out to the epoch of reionization, at which point a plateau appears in the CoDa~II patchy and non-patchy reionization cases, where the globally-averaged hydrogen ionized fraction drops below unity at $z\gtrsim6$ (as shown in Fig.~\ref{fig:ionization}). We also plot the results from \protect\citet{Jaroszynski_2019} as a comparison (black points) found using the data from the Illustris simulation. A slight difference between the results in \protect\citet{Jaroszynski_2019} and the two results we construct for the mean arises because we include a period of helium reionization at redshift $z=3$.}
    \label{fig:DMmean}
\end{figure}

\begin{figure*}
    \centering
    \includegraphics[width=\textwidth]{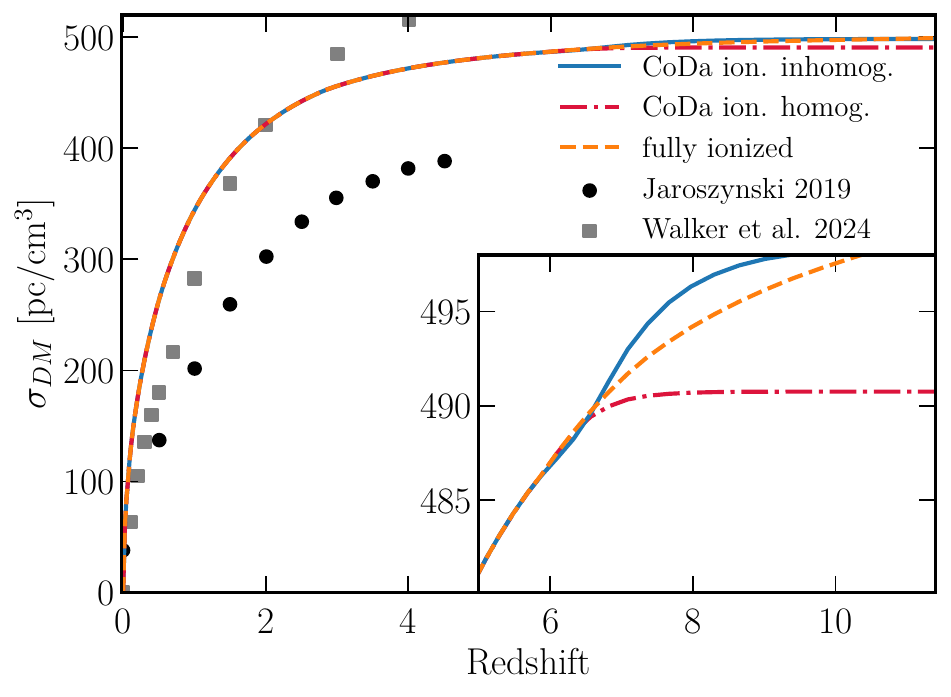}
    \caption{Standard deviation of dispersion measures modeled from CoDa~II/CoDa~II--Dark Matter simulations. Blue solid, orange dashed, and red dash-dotted curves represent the same set of dispersion measures as in Fig.~\ref{fig:DMmean}, but here the standard deviation rather than mean is plotted. The large difference between the standard deviation in the three results we construct and those from \protect\citet{Jaroszynski_2019} seems to arise primarily from simulation differences between the CoDa~II--Dark Matter and Illustris-3 simulations as well as differences in the analyses used in this work and that used in \protect\citet{Jaroszynski_2019} (black dots). The standard deviation constructed from Illustris-TNG data \protect\citep{Walker2023} (gray squares) is more consistent with the results we find in the present work.
    The inset figure zooms in on the time period around reionization. The most notable difference here is the increase in the standard deviation of the dispersion measures constructed using CoDa~II (blue solid curve, including the effects of a spatially inhomogeneous ionization fraction that on average begins to drop below $x_{HII}=1$ at $z\gtrsim 6$), 
    relative to the dispersion measures constructed from the CoDa~II simulation with the spatially averaged (but redshift dependent) ionization fraction (red dash-dotted curve), as well as to the dispersion measures constructed using the CoDa~II simulation where the hydrogen ionization fraction has artificially been set to 1 throughout (orange dashed curve, and consequently no patchiness). The relative difference between the two cases of redshift dependent ionization fraction, with (blue solid curve) or without (red dash-dotted curve) patchiness, reaches $\mathcal{O}(1\,\mathrm{per cent})$ of the total standard deviation at $z \sim 8$. 
    }
    \label{fig:DMstds}
\end{figure*}

In this section we discuss the results of our simulations for FRBs in the redshift range $0<z<12$.
In Fig.~\ref{fig:DMmean}, 
we show the mean of the dispersion measures calculated from the column densities constructed above, and in Fig.~\ref{fig:DMstds}, we show the standard deviation. In both figures,
the solid blue curve shows the results from the FRB dispersion measures calculated using the CoDa~II and CoDa~II--Dark Matter simulations (using CoDa~II--Dark Matter for $0<z<5.8$ and CoDa~II for all higher redshifts). As a comparison, the red dot-dashed curves show the results of dispersion measures constructed again using the CoDa~II and CoDa~II--Dark Matter simulations, but where we have averaged the ionization fraction $x_{HII}$ over space in each of the CoDa~II snapshots. As a result, the solid blue curve shows results that include inhomogenity (patchiness) of the ionization and the red dot-dashed curve shows results that assume homogeneous reionization and therefore include inhomogeneities only in the baryon density, and by comparing the two, we can infer the impact of the ionization inhomogeneities on the observable mean and standard deviation of the dispersion measures.  
As an addition comparison,
the orange dashed curve shows results from the CoDa~II--Dark Matter simulations alone.  The latter simulations assume that the universe is fully ionized with hydrogen ionization fraction $x_{HII}=1$ at all redshifts considered
\footnote{We have also studied the case of CoDa~II with $x_{HII} = 1$ everywhere to be able to compare the impact of CoDa~II--Dark Matter vs. CoDa~II baryon densities, and found negligible differences compared to the difference due to the effect of the different ionizations shown in the figures.}.

The dominant feature that can be seen when comparing the means of the dispersion measures is the plateau that occurs when full ionization is not assumed. This is expected naturally because the dispersion measure is sensitive only to the density of charged particles, so when the universe is largely neutral the dispersion measure does not increase noticeably. The dispersion measure only begins to vary once appreciable reionization occurs, allowing the location of the plateau to be a useful measure of when the EOR occurs. Using the CoDa simulation results, the dispersion measure of this plateau is approximately $\mathcal{DM}_{max} = 5900$.

Further, we plot results (black points) from the work of \citet{Jaroszynski_2019} found using data from the
Illustris-3 simulation to a redshift of $z=5$.   
Comparing the mean dispersion measure curves from our work (solid blue and dashed orange curves) to the results of \citet{Jaroszynski_2019}  also shows a small discrepancy that is to be expected. One of the assumptions we made was that helium reionizes from He II to He III at $z=3$. At earlier times ($z>3$), the number of free electrons contributed by each helium atom was assumed lower (since ionized H was accompanied by singly-ionized He then), so our dispersion measure was likewise reduced compared with its value if He were assumed to be doubly-ionized wherever H was ionized. The results in \citet{Jaroszynski_2019} assume He was doubly-ionized over the entire range of redshifts, including $z>3$, so they show a slight increase relative to our results.

In addition to the mean of the dispersion measure, we can look to the standard deviation of the dispersion measure to offer further information about the spatial inhomogeneity of the density and ionization fraction fields during the EOR. The paths that different FRB signals take travel through regions with differing densities of electrons. As a result, the standard deviation of dispersion measures includes a factor due to the inhomogeneity in density. However, there is an additional contribution to the standard deviation of FRBs emitted at redshifts prior to the end of the EOR due to FRB signals passing through inhomogeneously ionized regions of space. With the CoDa~II data, we can artificially homogenize this ionization field, allowing us to estimate the role the patchiness of reionization takes. We expect that the standard deviation of the distribution of dispersion measures out to a given redshift will be higher when we consider the effects of patchy or inhomogeneous reionization. In Fig.~\ref{fig:DMstds}, we can see this by comparing the blue solid curve to the red dot-dashed curve. The blue solid curve shows the standard deviation we calculate when including the effects of patchy reionization from the CoDa~II simulation, as a function of the redshift of a source FRB, while the red dot-dashed curve shows standard deviation when we apply the spatially averaged but time-evolving hydrogen ionization fraction to the CoDa~II density fields. That is, in both cases, the electron density fields are the same, and the average ionization fraction across the box is the same at every redshift, but the blue solid (inhomogenous) case allows patches within the simulation volume to have different hydrogen ionization fractions, while the red dot-dashed (homogeneous) case does not.

An important feature to observe in both the inhomogeneous and homogenous CoDa~II standard deviations is the plateau that is reached at redshifts higher than the end of the EOR. This plateau can be explained similarly to the plateau in the mean of the dispersion measure discussed above: namely that when the universe is essentially entirely neutral, it does not contribute to the dispersion measure. 
However, as illustrated in the inset of Fig.~\ref{fig:DMstds}, while the standard deviation of the homogeneous case reaches a plateau at the same redshift as the mean, the standard deviation in the inhomogeneous case only does so at a higher redshift. Furthermore,  the plateau in the standard deviation of dispersion measures in the homogeneous case is higher than in the inhomogeneous case, by a factor of approximately 1 per cent. These two effects, the higher redshift at which the plateau is reached and the increase in the standard deviation at that plateau, offer clues as to the nature of reionization.
For example, if the patches are typically small but numerous, the effect of the patchiness will be similar across most lines of sight, and there will be a smaller difference between the standard deviations of the homogeneous and inhomogeneous cases. Conversely, if reionizing patches are sparse and consequently grow to large size, different lines of sight can experience dramatically different ionization fractions, and the difference in the standard deviations will be larger.

In addition, in Fig.~\ref{fig:DMstds}, we provide two additional comparisons to the two cases that use the CoDa~II ionization history. The orange dashed curve shows the standard deviation of dispersion measures if the gas in the universe is fully ionized to redshifts of at least $z=12$. To construct this curve, we use the CoDa~II densities, but set the hydrogen ionization fraction artificially to be $x_{HII} = 1$ everywhere. In addition, we plot the standard deviations from \citep{Jaroszynski_2019} as black points. Jaroszynski's analysis extends only to a redshift $z=5$, so does not offer any comparison during the EOR, but provides a point of comparison at lower redshifts. Of particular note, the standard deviation we find is larger than the standard deviations found by Jaroszynski by a factor of approximately $5/4$. We discuss this difference in more detail in Appendix~\ref{appendix}. We propose three partial explanations for this difference, but leave it to future work to determine which, if any, of these effects dominate. The CoDa~II--Dark Matter simulation has a higher mass resolution than the Illustris-3 simulation that \citet{Jaroszynski_2019} analysed. This higher resolution allows many low-mass structures to form. Because the standard deviation of the dispersion measure is sensitive to the high-dispersion measure tail of the distribution, filling out this tail with more dispersion measures from lines of sight that pass through small overdensities could shift the standard deviation to higher values. Alternatively, in works like \citep{McQuinn_2013}, the standard deviation of dispersion measures from dark matter-only models were systematically higher than the standard deviations from simulations including baryonic effects. The Illustris simulation includes those baryonic effects, while the CoDa~II--Dark Matter simulation we employ at redshifts $z<5.8$ is a dark matter-only simulation. This difference could also contribute to the observed difference between the standard deviations we find in this work and those from \citep{Jaroszynski_2019}. We note, however, that direct comparisons of the CoDa~II--Dark Matter simulation at higher redshifts with the CoDa~II simulation that includes baryonic effects show nearly identical standard deviations. Finally, authors including \citet{Park2018} have argued that the first generation of the Illustris simulation, including Illustris-3, includes AGN feedback that over-suppresses structure growth at low redshifts, leading to an overall reduction in the variance of contributions to the dispersion measure from low redshifts. Analyses of other, more recent simulations that do not have this exaggerated AGN feedback, including Illustris-TNG, bear out standard deviations that are comparable to or higher than the ones we find using the CoDa simulation \citep{Walker2023}.

Our primary interest in this paper is to explore the epoch of reionization using FRB dispersion measures, and this is achieved predominantly through differences in the standard deviation of dispersion measures from FRBs at prior to or during the EOR. From the results in Fig.~\ref{fig:DMstds}, based on the CoDa~II simulation, we estimate the size of the effect of patchy reionization on the standard deviation of dispersion measures to be approximately 1 per cent, for FRBs at $z\gtrsim 8$. 
From this, we can make a rough estimate of how challenging it would be to observe an effect of this size, while we leave a more comprehensive analysis to future work. First, the results we present here show the statistics of the dispersion measure as a function of redshift, which would only be directly observable if the redshifts of FRB sources could be identified. As we look to higher redshifts, it is likely to become more challenging to precisely identify a redshift for any individual FRB source. As such, observationally, the most practical statistic to look at is the distribution of FRB dispersion measures. An example of a semi-analytic calculation of this distribution can be seen in Fig.~3 of \citep{Beniamini2021}. To explore the EOR, the most relevant feature in the distribution of FRB dispersion measures is a buildup of FRBs with the maximal dispersion measure, which occurs because all FRB signals emitted before the EOR will have comparable dispersion measures. The dispersion measure at which this peak occurs will be related to what appears in our analysis as a plateau, so from the CoDa~II simulation we would expect the peak to occur around $5900 ~\mathrm{pc\,cm^{-3}}$. The shape of this peak, particularly the width about the mean, will be related to the standard deviation of the dispersion measures for FRB signals emitted prior to the end of the EOR. With this in mind, ignoring other sources of error, we can estimate that in order to detect a 1 per cent change in the standard deviation of the dispersion measure, we would need at least 1 per cent resolution on the width about this peak. Since standard error scales as $1/\sqrt{N}$ where $N$ here is the number of FRBs observed, we anticipate needing to observe $\mathcal{O}(10^{4})$ FRBs 
emitted prior to the end of the EOR, in order to determine the standard deviation to high enough precision to compare with the results we show here. \citet{Beniamini2021} report an estimate of approximately $10^4$ FRBs occur per day and estimate that approximately 0.1 per cent of detectable FRBs are produced prior to the end of reionization. Independent of other sources of uncertainty, an observation time of three years could be sufficient to compare to our results.

\begin{figure*}
    \centering
    \includegraphics[width=\textwidth]{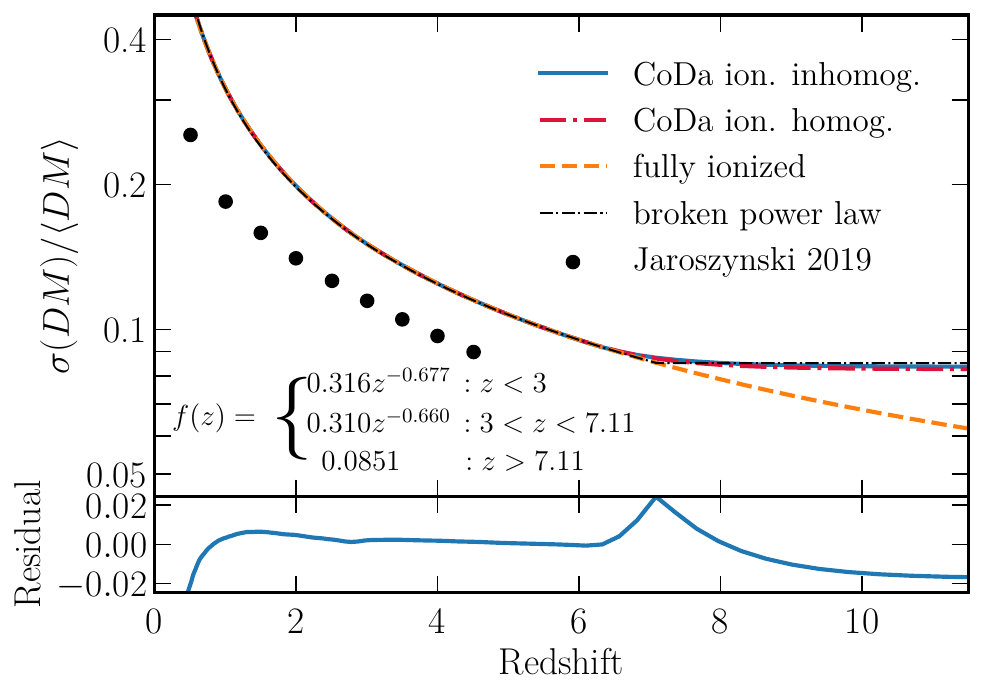}
    \caption{Coefficient of variance of dispersion measures. We plot the ratio of standard deviation to mean for the scenarios described in Fig.~\ref{fig:DMmean} on a log-linear plot. The case where the gas is fully ionized throughout the redshift range $0<z<12$ (orange dashed curve) fits a power law, while the dispersion measures constructed from the combination of the CoDa~II and CoDa~II--Dark Matter simulations (blue solid curve) approximately fit a broken power law (dot-dashed black curve, with fit $f(z)$ as described inside the plot). Residuals between solid blue curve and the broken power law that best fits that curve (black dash-dotted curve) are also shown. While the difference is generally small, two peaks occur: one at the kink in the broken power law and the other at low redshifts for which the power law becomes a less reliable fit. }
    \label{fig:norm+resid}
\end{figure*}

Finally, we look at the coefficient of variance, the standard deviation of dispersion measures along different lines of sight normalized to the mean dispersion measure, which we present in Fig.~\ref{fig:norm+resid}. This ratio is well-modeled by a power law in the CoDa~II--Dark Matter case, and by a broken power law for the combined CoDa~II and CoDa~II--Dark Matter case. At lower redshifts, approximately $z \lesssim 0.5$, the ratio $\sigma(DM)/\langle DM\rangle$ rapidly decreases. 
The broken power law that best fits the coefficient of variance is
\begin{equation}
    \frac{\sigma(DM)}{\langle DM\rangle} \approx 
    \begin{cases}
        0.316 z^{-0.677} & z<3\\
        0.310 z^{-0.660} & 3<z<7.11\\
        0.0851 & z\geq 7.11
    \end{cases}.
    \label{powerlaw}
\end{equation}
Here, the first break is fixed at redshift $z=3$, corresponding to the redshift at which we set the reionization of helium. The redshift of the second break is fitted, but corresponds to the redshift at which the intergalactic medium is half-ionized.
As we will discuss below, power laws of this form may be particularly useful in distinguishing between different models of reionization, and the one described here offers a baseline to compare against.

To confirm the quality of the fit, we also plot the residuals between the coefficient of variance from the constructed electron densities from the CoDa~II and CoDa~II--Dark Matter simulations and the best-fitting broken power law above in Fig.~\ref{fig:norm+resid}.  
The (relative) difference is fairly small, mostly on the order of 0.5 per cent
over all redshifts that we consider. However, there are two dominant features where the residual increases by a factor of a few. One in the redshift range $0<z<2$, and one around redshift $z=7$. Both of these peaks hint at situations in which modeling the coefficient of variance as a broken power law is not accurate to the smooth nature of the data. The feature around redshift $z=7$ arises because we are modeling a smooth plateau as a sharp kink. At lower redshifts, the coefficient of variance shows a more pronounced deviation from the power law. This may reflect a physical meaningful reduction in the coefficient of variance relative to a pure power law at redshifts $z<0.5$.

\section{Effects of Alternate Global Reionization History}
\label{toys}
Throughout this work, we use the results of the CoDa~II simulation as the basis for our ionization fraction and density fields. However, all of these results are subject to modeling uncertainties (e.g. star formation efficiency), so in this section, we explore the extent to which one such variability can affect the results we discuss above. Namely, we alter the globally-averaged reionization history (specifically, the redshift dependence of the spatially-averaged ionized fraction $\langle x_{HII}\rangle$). To do this analysis, we first calculate the spatially-averaged ionization fraction as a function of redshift from the CoDa~II data. We then compute a fluctuating electron density field by multiplying the inhomogeneous matter density field from the CoDa~II--Dark Matter simulation by this spatially-averaged ionization fraction ($(1+\delta_H)\langle x_{HII}\rangle$). We can then model different ionization histories by modifying the redshift function of ionization fraction from the CoDa~II ionization history. However, because we are spatially averaging the ionization fraction, 
we cannot consider in this section the effects of inhomogeneities in the ionization fraction on the standard deviation, as we did in the previous section. 
As such, we expect that a similar $\sim 1$ per cent increase to the standard deviation in dispersion measures that we previously found when including inhomogeneity would be observed in addition to any differences in the standard deviation that arise from the alternate histories considered below.

\begin{figure}
    \centering
    \includegraphics[width=\columnwidth]{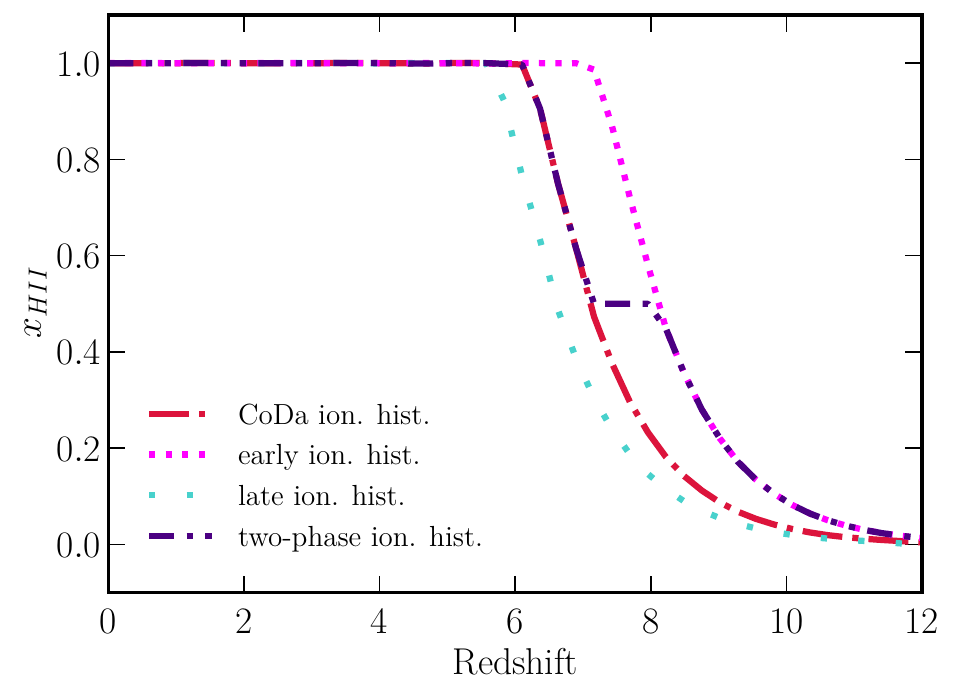}
    \caption{Average hydrogen ionization histories ($\langle{x_{HII}\rangle}(z)$) in each of the reionization histories we consider. 
    As in figure~\ref{fig:ionization}, each curve represents the spatially averaged ionization fraction as a function of redshift. The red dash-dotted curve shows the hydrogen ionization history of the CoDa~II simulation without modification, which shows the effects of a homogeneous non-patchy reionization. The magenta densely dotted (teal sparsely dotted) curve shows the ionization history if the universe ionized earlier (later) than in the CoDa~II simulation, so that the end of reionization is approximately $z=7.1$ ($z=5.6$). The indigo dash-dot-dotted curve shows the ionization history if reionization occurs in two phases, where 50 per cent of the universe is ionized in each phase, but ionization fraction plateaus between the two phases. 
    }
    \label{fig:ion_comp}
\end{figure}

We consider four different homogeneous CoDa~II ionization histories: three new ones in addition to the previously studied case of `CoDa ion. homog.' (homogeneous but time-dependent reionization).
The ionization fractions in these four cases are plotted as a function of redshift in Fig.~\ref{fig:ion_comp}. 
We note that all models are the same for redshifts from today up to the end of reionization ($z=6.1$ in the canonical case), with the differences playing a role at redshifts higher than that. These four modifications can be summarized as follows.

\begin{itemize}
    \item \textbf{CoDa~II ionization history} This model `CoDa ion. homog' (studied earlier in the paper) uses the mean hydrogen ionization fraction within each CoDa~II simulation box to determine $x_{HII}(z)$. As such, it largely reproduces the CoDa~II results as if there were no patchy reionization and the entire universe reionized at the same rate everywhere. This is, therefore, perhaps the most direct comparison to the patchy reionization results.
    \item \textbf{Early reionization} In this model, we use a hydrogen ionization fraction that is shifted in redshift: $x_{HII}(z) = x_{HII, CoDaII}(z-1)$. Effectively, this amounts to shifting the reionization to an earlier time, so that instead of ending at $z = 6.1$ it ends at $z=7.1$. The hydrogen ionization history is still the same as the hydrogen ionization history from the CoDa~II simulation, but shifted in redshift.
    \item \textbf{Late reionization} Similar to the early reionization model, we use a hydrogen ionization fraction that is shifted in redshift, but now to lower redshifts: $x_{HII}(z) = x_{HII,CoDaII}(z+0.5)$. By shifting the ionization history to slightly later redshifts, we approximate more recent observational findings of a later end of the EOR. This model has the EOR ending at approximately $z=5.6$, and the universe being half-ionized at approximately $z\approx 6.5$.
    \item \textbf{Two-phase ionization history} In this model, the hydrogen ionization history is split into two phases. Again based on the hydrogen ionization history from the CoDa~II simulation, we created two phases of reionization. When $x_{HII}<0.5$, we use the CoDa~II hydrogen ionization history directly, but when $x_{HII}>0.5$, we use the shifted hydrogen ionization history: $x_{HII}(z) = x_{HII, CoDaII}(z-1)$. Effectively, in this scenario the universe ionizes until half ionized and then experiences a pause, before ionizing completely some time later. 
    An example of a physical effect which results in the two-phase global reionization history is the inclusion of minihaloes--haloes with mass $M\lesssim 10^8 M_\odot$--as sources of star formation \citep{Ahn2012,Shapiro2021}. Because minihaloes are the earliest haloes to form, they are also the first sources of UV starlight, and they initially dominate reionization.  By the time a significant fraction of the volume is reionized, however (e.g. as high as $\sim 0.2 - 0.3$), the ability of minihaloes to form stars, involving $H_2$ formation and cooling, is suppressed by the rising UV background of $H_2$-dissociating starlight in the Lyman-Werner bands, and the contribution of reionizing photons from minihaloes begins to diminish.   At this point, reionization stalls until more massive atomic cooling haloes can form in abundance to take over the process and finish reionization. The combination of these two periods of reionization, the first dominated by minihaloes and the second dominated by atomic cooling haloes, produces an early tail to the ionization history, similar to the model we show here.
    
\end{itemize}

\begin{figure*}
    \centering
    \includegraphics[width=0.45\textwidth]{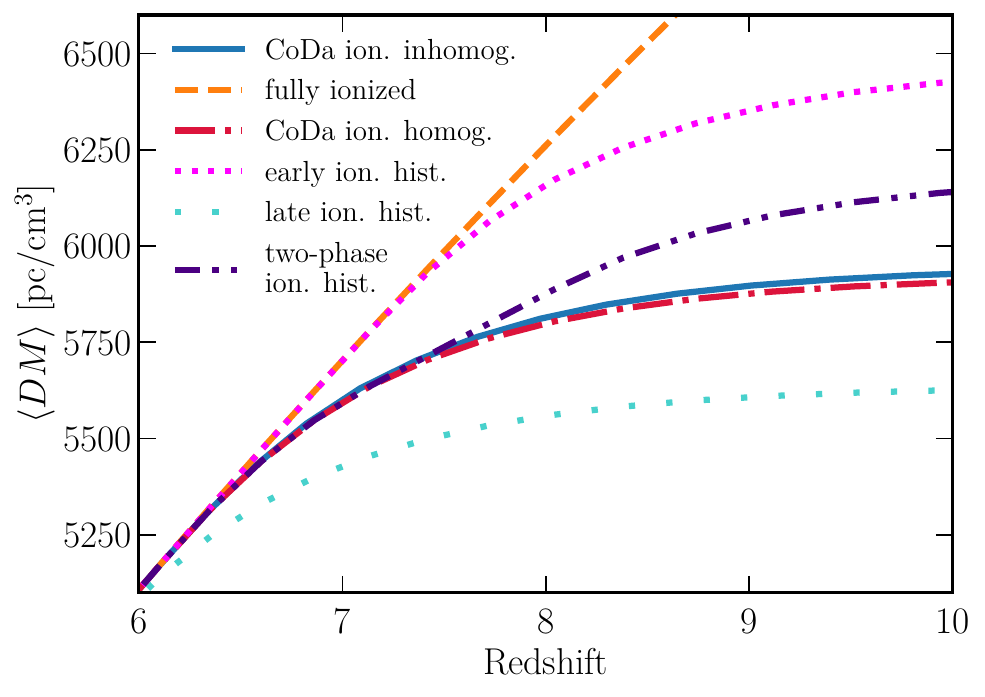}
    \includegraphics[width=0.45\textwidth]{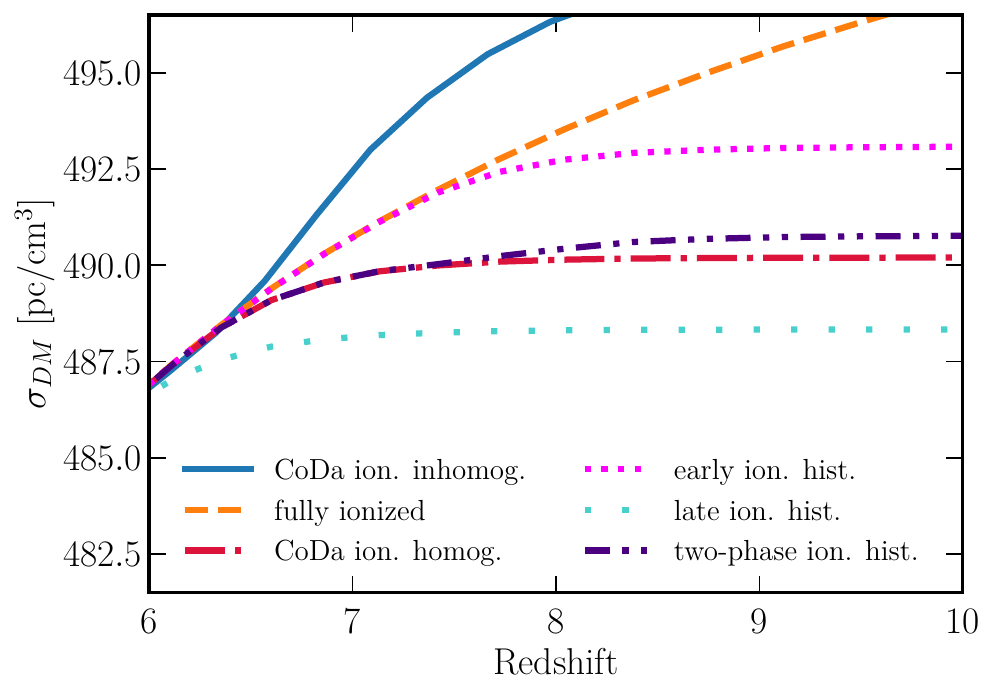}
    \caption{Comparison of dispersion measure means (left) and standard deviations (right) during the epoch of reionization from various ionization histories. As a reference, the mean and standard deviation of dispersion measures assuming a fixed hydrogen ionization fraction of $x_{HII}=1$ is plotted as the orange dashed curve. The blue solid curve represents the dispersion measures derived from the CoDa~II ionization history, including patchy reionization. The remaining curves each involve the CoDa~II ionization history, averaged over the simulation volume, and modified in various ways, described in the main text. In each case, both the mean and standard deviation have different plateau values that occur at redshifts prior to the end of reionization. For the mean dispersion measures shown in the left panel, one can see that the case of full ionization ($x_{HII}=1$, orange dashed cure) is an upper bound on all of the models which include reionization, and that the plateau in the mean dispersion measure increases the earlier the EoR ends and for longer EoR durations. 
    Likewise, the orange dashed curve is an upper bound on all of the standard deviations of dispersion measures calculated with spatially homogeneous ionization fraction. However, once inhomogeneities in the ionization are taken into account (blue solid curve), the standard deviation can increase above the orange curve. Furthermore, while different ionization histories predict different standard deviations of dispersion measures, the models we consider all predict standard deviations within $\mathcal{O}$(1 per cent) of one another. This can be compared to the difference between the standard deviations calculated from the CoDa~II simulation with homogeneous (red dash-dotted curve) and inhomogeneous (blue solid curve) ionization fractions, which is also $\mathcal{O}$(1 per cent).  
    }
    \label{fig:toy-models}
\end{figure*}

In Fig.~\ref{fig:toy-models}, we compare the mean and standard deviation of the dispersion measures calculated using these alternate ionization histories to the dispersion measures using the CoDa~II--Dark Matter (fully ionized) and combined CoDa~II--Dark Matter and CoDa~II (patchy reionization) models. As expected the mean dispersion measure is very similar between the patchy reionization history and the model based on the unmodified CoDa~II ionization history `CoDa ion. homog' with time-dependent ionization but without patchiness.  The other models (all with spatially homogeneous ionization fractions), however, diverge from this model in different ways. In particular, the early ionization model plateaus at a higher redshift than the CoDa~II model, and consequently reaches a higher mean dispersion measure. The two phase reionization model follows the plateauing of the CoDa~II reionization, but then diverges and continues increasing until plateauing at a lower dispersion measure compared to the early ionization history. In general, each of these mean dispersion measures follow the ionization fraction history of the model associated with them, as we would expect.

The standard deviations of the dispersion measures from each of these models likewise have features tied to the ionization history. A particularly notable feature is the distinctly large difference between the standard deviations from the patchy reionization history and the toy model based on the unmodified CoDa~II ionization history. Comparison of these two curves offers perhaps the most direct way to explore the effect of the patchy reionization versus the same reionization history without the patchiness. The difference in the maximum standard deviations achieved by these two models is, in fact, larger than the differences between any of our other reionization history models. This implies that the patchiness of the reionization specifically will produce a distinct pattern in the standard deviation history of FRBs. If sufficient numbers of FRBs can be detected at redshifts prior to the EOR, and their dispersion measures accurately extracted, the standard deviation of the dispersion measure could allow a measure of the typical spatial uniformity of the patches reionization occurs in.

\begin{figure}
    \centering
    \includegraphics[width=\columnwidth]{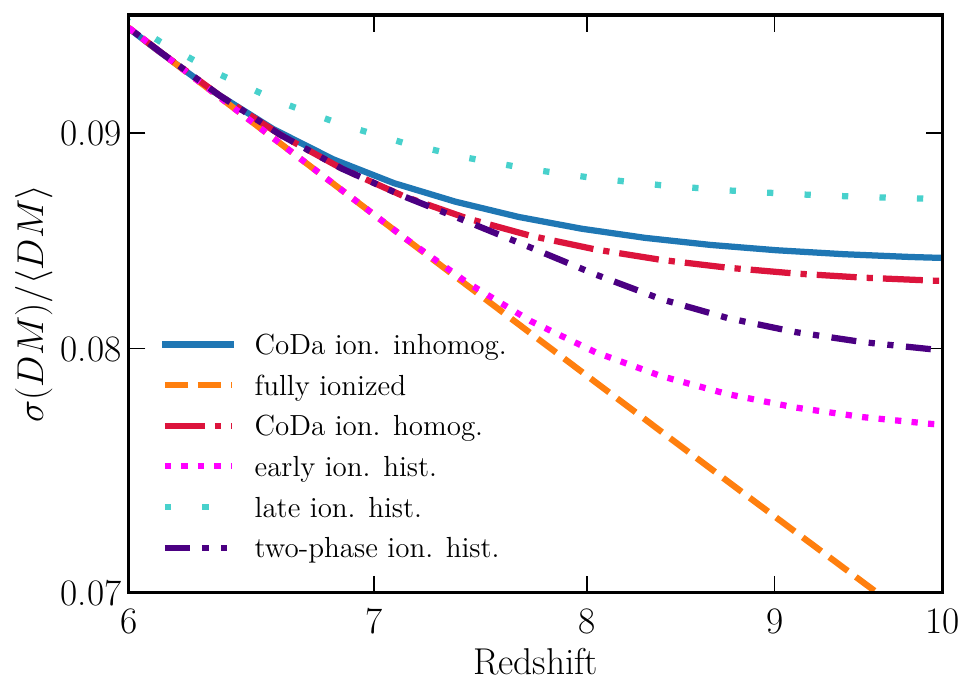}
    \caption{Comparison of the coefficient of variance of dispersion measures modeled using various ionization histories. Curves have the same meaning as in Fig.~\ref{fig:toy-models}. While the coefficient of variance from each model can be described by a common power law at lower redshifts than reionization, the constant value at which the coefficient of variances plateaus is different for each model.}
    \label{fig:toy-ratios}
\end{figure}

In addition to the two comparisons separately, in Fig.~\ref{fig:toy-ratios}, we compare the coefficients of variance as a function of redshift for the alternate reionization histories.  
At low redshifts, the coefficient of variance follows the power law fit in the previous section, but at higher redshifts it can offer a probe of the epoch of reionization that is complementary to the standard deviation and mean separately. Like the mean, the coefficient of variance is more sensitive to differences in the average ionization history than to whether the ionization is homogeneous or inhomogeneous. Combining these statistical quantities therefore offers a potential way to constrain the timing and history of reionization. For example, the coefficient of variance follows the power law identified above at low redshifts, and deviates from this power law at a point determined by the end of the EoR. Meanwhile, at high redshifts, the coefficient of variance is constant and deviates from this plateau value at a redshift related to the beginning of the EoR. If enough FRBs can be observed to span this entire redshift range, from beginning to end of EoR, the standard deviation, mean, and coefficient of variance may offer a means to probe the evolution of the EoR. Even if sufficient numbers of FRBs cannot be observed at all redshifts in this range, FRBs in multiple parts of this range could probe different aspects of the EoR.

\section{Conclusion}
\label{conc}
Using the results of the CoDa~II simulations, we have predicted the mean and standard deviation of the dispersion measures of FRBs, as a function of the source redshift of those FRBs. In doing so, we observe and quantify the theorized plateaus in the mean dispersion measure and in the standard deviation around that mean at redshifts higher than the end of the EOR. 
We present these results in a way that can be most directly applied if FRBs are localized to a well-defined redshifts. However, even if well-constrained redshift information cannot be obtained, the plateaus we describe would lead to a peak in the distribution of dispersion measures. We predict that the plateau dispersion measure, and consequently the dispersion measure at which this peak occurs, is $\mathcal{DM}_{peak} \approx 5900~\mathrm{pc\,cm^{-3}}$. The value of this peak dispersion measure can be related to the redshift at which reionization ends, where a later end of the EOR leads to a lower peak dispersion measure, and to the duration of the EOR, where a longer period of reionization leads to a higher peak dispersion measure.

Additionally, we predict the standard deviation of dispersion measures at redshifts $z\gtrsim 8$, which will be related to the spread about the peak dispersion measure, to have a value of $\sigma_{DM} \approx 497 ~\mathrm{pc\,cm^{-3}}$, approximately 10 per cent of the mean value. This standard deviation is related to the evolution of inhomogeneities in both the ionized gas and overall gas density. 
We quantify how much of this standard deviation can be attributed to the inhomogeneity in ionization fraction due to the patchy nature of reionization by comparing the standard deviation with the inhomogenous ionization to the standard deviation with the same ionization history but a spatially uniform ionization fraction.   

One of our main results can be found in the inset of Figure 4 as described in Section 4, where we showed that evidence for patchy reionization may be found at the $\mathcal{O}$(1 per cent) level once FRBs are found at $ z \gtrsim 8$.   
Specifically we find that the patchiness of reionization contributes approximately a 1 per cent increase to the standard deviation that would be observed from FRBs with redshift $z \gtrsim 8$.
A back-of-the-envelope calculation suggests that this could optimistically be resolved if $\mathcal{O}(10^4 - 10^5)$ FRBs emitted prior to the end of the EOR can be observed. Using estimates for the rate at which FRBs are produced, it may take as little as 3 years for this number of FRBs to be produced, suggesting that even this 1 per cent effect is within observational reach. 

We also compute the coefficient of variance, $\sigma_{DM}/\langle \mathcal{DM} \rangle$, as a function of redshift, and find that it is well fit to a broken power law, as in equation~\eqref{powerlaw}. For redshifts $z\leq 3$, when helium is fully ionized, the coefficient of variance can be represented by the power law $\frac{\sigma_{DM}}{\langle \mathcal{DM} \rangle} = 0.316 z ^{-0.677}$. For redshifts $3<z\leq 7.11$, $\frac{\sigma_{DM}}{\langle \mathcal{DM} \rangle} = 0.310 z ^{-0.660}$. And prior to the end of reionization, the coefficient of variance also plateaus at a value of $\frac{\sigma_{DM}}{\langle \mathcal{DM} \rangle} = 0.0851$. We hope that power laws like these can be used in future semi-analytical works to estimate the distribution of dispersion measures.

Finally, we explore how our results might shift if the ionization history is not identical to the one predicted from the CoDa~II simulation. We consider three alternative models of the global ionization fraction, mimicking universes that see reionization start and end earlier than in the CoDa model, start and end later than in the CoDa model, and start earlier but end at the same time as the CoDa model. Each of these alternate ionization histories produces slightly different peak values in the mean of the dispersion measures, standard deviations about those peak values, and value of the coefficient of variance at redshifts higher than the end of the EOR. Therefore, combining measurements of these three properties offers a direct probe of the global ionization history.

\section*{Acknowledgements}

TD and PRS are grateful for the support of the  NSF Graduate Research Fellowship Program under Grant No. DGE-1610403. This research used resources of the Oak Ridge Leadership Computing Facility (OLCF) at Oak Ridge National Lab, which is supported by the Office of Science of the US Dept. of Energy (DOE) under Contract No. DE-AC05-00OR22725. CoDa II was performed on Titan at OLCF under DOE INCITE 2016 award to Project AST031. Data analysis was performed using computing resources from NSF XSEDE grant TG-AST090005 and the Texas Advanced Computing Center at the University of Texas at Austin. PRS also acknowledges support from NASA under Grant No. 80NSSC22K175.
KF is Jeff \& Gail Kodosky Endowed Chair in Physics at the University of Texas at Austin, and KF and JJZ are grateful for support via this Chair. KF and JJZ acknowledge support by the U.S. Department of Energy, Office of Science, Office of
High Energy Physics program under Award Number DE-SC-0022021 as well as support from the Swedish Research Council (Contract No. 638-2013-8993). PB is supported by a grant (no. 2020747) from the United States-Israel Binational Science Foundation (BSF), Jerusalem, Israel, by a grant (no. 1649/23) from the Israel Science Foundation and by a grant (no. 80NSSC 24K0770) from the NASA astrophysics theory program.
PK is supported by an NSF grant AST-2009619, and a NASA grant 80NSSC24K0770 (PK and PB).
KA is supported by NRF-2021R1A2C1095136 and RS-2022-00197685.

\section*{Data Availability}
The data underlying this article are a product of The Cosmic Dawn (``CoDa'') Project. The derived data generated in this research will be shared upon reasonable request to the corresponding author.

\bibliographystyle{mnras}
\bibliography{FRB}

\appendix
\section{Additional Methods}
\label{moremethods}
In the main text, we outline the method we use to calculate the dispersion measure using the CoDa simulations. In this section we expand upon some of the details in that discussion. Our goal is to approximate equation~\eqref{cosmic dispersion measure} by using overdensity $\delta_H$ and ionization fraction $x_{HII}$ fields that are defined only at discrete redshifts.

The first step of this analysis is to determine the factors $x_e(z_i)$ and $\delta_H (z_i)$ for appropriate values of $z_i$ along different lines of sight. If we imagine tiling simulation volumes along the lines of sight, then the redshifts $z_i$ should be such that $t(z_i) - t(z_{i-1}) = L_b/c$, or that the time it takes light to cross the box should ideally be the same as the time interval at which we evaluate $x_e$ and $\delta_H$. The CoDa~II and CoDa~II--Dark Matter simulations were not designed to have this requirement, so either have simulated snapshots either more closely spaced than the ideal (in the CoDa~II simulation) or more sparsely spaced than the ideal (in the CoDa~II--Dark Matter simulation), as depicted in Fig.~\ref{fig:stitching}. We therefore process these two simulations differently so as to achieve the desired spacing in time/redshift.

In the first case, the time separation between subsequent snapshots is smaller than the target (the CoDa~II simulation).In this case, a photon traveling across a distance equal to the simulation box will experience multiple snapshots before fully crossing the box. Equivalently, a photon will move through only a fraction of the box described by one snapshot before the time of the succeeding snapshot. Therefore, when we calculate the contribution to the dispersion measure from one box-length segment of a line of sight, we incorporate the updated density and ionization information that a photon traveling through the box would see.

The second case is the opposite extreme, where the time between simulation snapshots is longer than the target (the CoDa~II--Dark Matter simulation). As a result, we seek to construct distributions of box-length contributions to the dispersion measure for redshifts between two simulation snapshots. For a given redshift, the distribution of the contributions to the dispersion measure is generated from the simulation snapshots immediately before and after that redshift, such that statistically the contributions to the dispersion measure are a linear interpolation between the contributions that would be calculated from those simulation snapshots. Effectively, this approach allows us to approximate changes to the density and ionization fraction, so that the universe `evolves' at least once per box-length that a photon travels.

Our goal in constructing these distributions is only to achieve something that is statistically representative, rather than physically accurate. Consider a redshift $z$ between two simulation redshifts $z_l$ and $z_r$ such that $z_l<z<z_r$. We can construct a distribution from each of the simulation redshifts, by considering the contribution to the dispersion measure along each line of sight as independent random elements of this distribution. Then, to calculate the distribution for the redshift $z$, we can average between the simulation distributions at $z_l$ and $z_r$. That is, we create a new distribution by selecting 
\begin{equation}
N_{l} = \frac{z_{r} - z}{z_{r} - z_{l}}N
\end{equation}
elements from the simulation distribution at $z_l$. Here, $N$ is the total number of elements in the distribution we seek to create. Likewise, we also include $N_r=N-N_l$ elements randomly selected from the simulation distribution at $z_r$. While other approaches can be used to generate these distributions, this approach retains any skewness that the simulation distributions have, so long as a sufficient number of elements from each of the simulation distributions are chosen. See appendix~\ref{appendix} for an alternative approach, and how it compares to the one used throughout.

Once we have a distribution of contributions to the dispersion measure from each box-length segment, we combine them into a set of lines of sight. As mentioned in the main text, we do not use the full $3\times2048^2$ possible contributions from each segment, but select 2 million from each. We combine these randomly, without repeating. Effectively, this amounts to each line of sight following a photon path through a continuous stack of simulation volumes that are rotated and translated randomly, and independently of any other lines of sight. This approach could limit any correlations that might exist between nearby lines of sight over distance scales greater than the size of the simulation box. However, it is expected that there are very few correlations in density on scales greater than 100 Mpc, so we assume any limitation due to the randomization is negligible. Finally, segment length contributions are added together so that the total line of sight reaches a given redshift.

\section{Illustris Dataset}
\label{appendix}
Throughout this paper, we have used a methodology designed to compliment the Cosmic Dawn (CoDa)~II simulation, as well as the associated CoDa~II--Dark Matter simulation. To confirm that this approach is consistent with other methods in the literature, we attempted to match the results found in \citet{Jaroszynski_2019} based on the Illustris-3 simulation \citep{Nelson_2015}. While successfully able to reproduce these results, and to show marked similarity between the dispersion measures calculated using the Illustris and CoDa~II--Dark Matter simulations at low redshifts, we also identify a systematic difference between the approach we develop above and the approach that matches the results in \citet{Jaroszynski_2019}, which may have significant repercussions.

Following \citet{Jaroszynski_2019}, the Illustris datasets we used were the Illustris-3 snapshots at redshifts $z=0.00$, 1.00, 2.00, 3.01, 4.01, and 5.00. However, these datasets store data differently than the CoDa~II simulations, with the Illustris datasets storing data about cells with variable volumes but a similar mass, while the CoDa~II and CoDa~II--Dark Matter datasets store data about fixed volume cells which consequently contain differing amounts of mass. For our analysis, we converted the Illustris dataset to the spatial grid utilized by CoDa~II and CoDa~II--Dark Matter by taking a simple sum of the dark matter and gas particle masses whose centres were located in each spatial grid subdivision. This simple summing procedure may have introduced inaccuracies to our reproduction of the Illustris dataset, but it is believed that these inaccuracies are small.

\begin{figure*}
    \centering
    \includegraphics[width=0.45\textwidth]{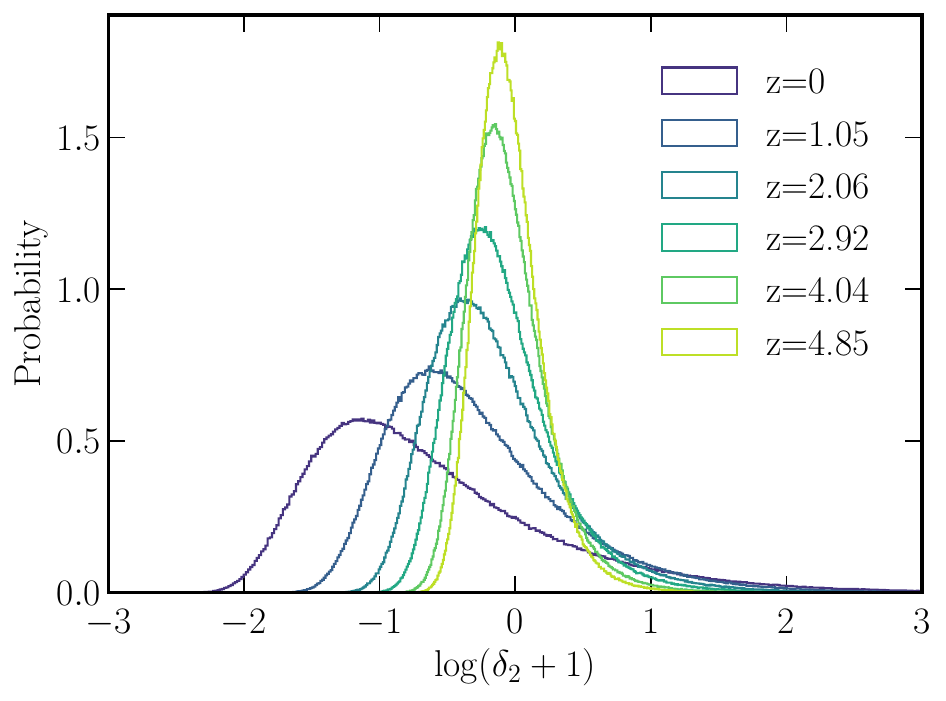}
    \includegraphics[width=0.45\textwidth]{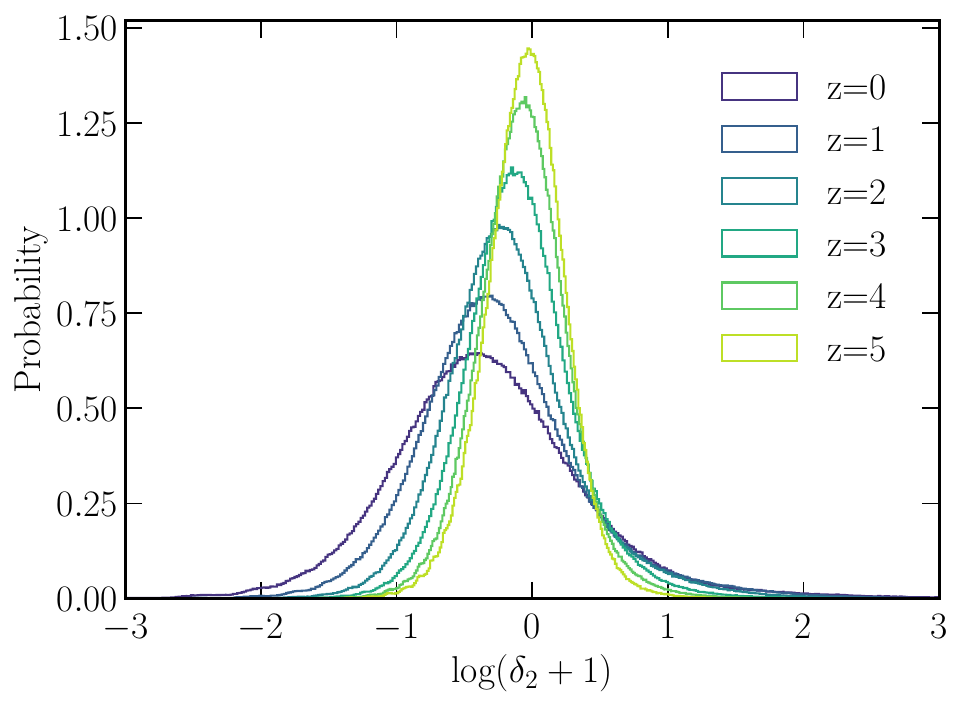}
    \caption{Comparison of the distribution of overdensities at low redshifts in the CoDa~II--Dark Matter and Illustris simulations. We compare the distribution of the logarithm of the overdensity $\log(1+\delta_2)$ from the CoDa~II--Dark Matter (left) and Illustris (right) simulation results at similar redshifts. While the Illustris overdensities are mostly log-normal even to low redshifts, the CoDa~II--Dark Matter overdensities show distinct skewness, reflecting a higher number of high-density haloes in the dark-matter-only evolution. In both simulations, higher redshifts show less skew than $z=0$.}
    \label{fig:distcomp}
\end{figure*}

Next, we use each snapshot to generate a distribution of the overdensity $\delta_H$ from the snapshot volume for both the Illustris and CoDa~II--Dark Matter simulations. Fig.~\ref{fig:distcomp} shows a representation of these distributions, specifically $(1+\delta_2)$, which is the mean of the quantity $(1+\delta_H)$ along a line through the snapshot volumes. As elsewhere in the paper, this distribution can be approximated well by taking only lines perpendicular to one of the faces of the simulation volume and considering 2 million randomly selected elements of the 12 million independent values at each redshift. The approach taken in \citet{Jaroszynski_2019} uses the same simplification, but allows for a larger number of statistically independent lines. This difference in the number of lines has no impact on the statistical results we find.

As described in appendix~\ref{moremethods}, we can use these simulation snapshots to construct a series of synthetic snapshots that entirely covers the range, so as to probe the dispersion measure out to any given redshift $z$. The methodology of \citet{Jaroszynski_2019} uses a similar approach, but with slight differences. In either case, given a target redshift $z$ bounded by redshifts $z_{left}$ and $z_{right}$, for which there are simulation snapshots, we calculate a factor
\begin{equation}
    x = \frac{z - z_{left}}{z_{right} - z_{left}},
\end{equation}
which provides a measure for how close the redshift $z$ is to $z_{left}$ relative to $z_{right}$. Then, in the approach we use throughout this paper, we randomly select $xN$ elements from the snapshot at redshift $z_{left}$ and $(1-x)N$ elements from the snapshot at redshift $z_{right}$, where $N$ is the total number of elements considered in the distribution we consider at each redshift. Throughout this paper, we have used $N=2000000$, but as long as this number is sufficiently large to be representative of the simulation snapshots, the value itself does not have a large impact on the results.

In order to compare against the approach used in \citet{Jaroszynski_2019}, we followed the methodology outlined there, although minor adjustments needed to be made to accommodate the additional conversion of the Illustris data to a fixed grid.
Beginning with the simulation snapshots, we generate a distribution $\delta(z_{snapshot})$, based on a histogram of the logarithm of the contribution to the dispersion measure from each snapshot. The histograms we use have 560 bins between $\log(DM/\langle DM\rangle) \in [-2.5, 3.1]$. It turns out that the results are highly sensitive to this range, with the bounds listed here being comparable to those used in \citet{Jaroszynski_2019}. In that work, it was reported that this range contained all densities $\delta$, while our analysis presented a small number of densities outside this range; it is probable that these outliers arise from the pre-processing that we do to conform the Illustris data to the spatial gridding of the CoDa~II simulation. In any case, with these distributions generated from each snapshot, we can construct the distributions associated with synthetic snapshots as 
\begin{equation}
    \delta(z) = x \delta(z_{left}) + (1-x) \delta(z_{right}).
\end{equation}
We then randomly select $N$ elements from this distribution.

\begin{figure*}
    \centering
    \includegraphics[width=0.45\textwidth]{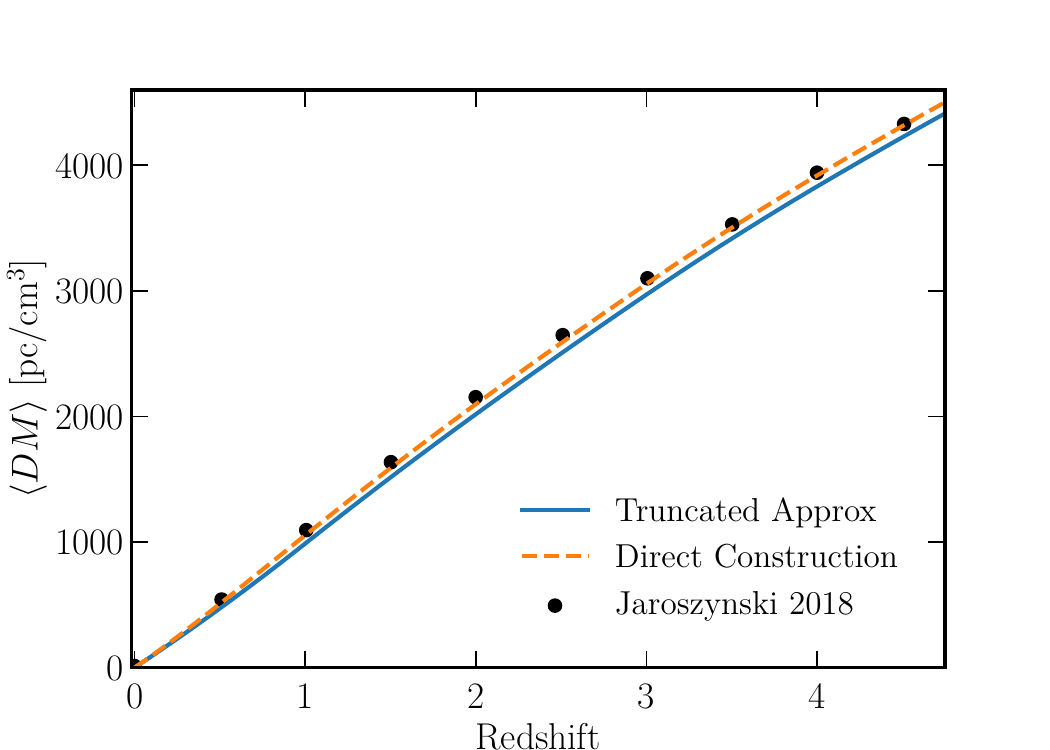}
    \includegraphics[width=0.45\textwidth]{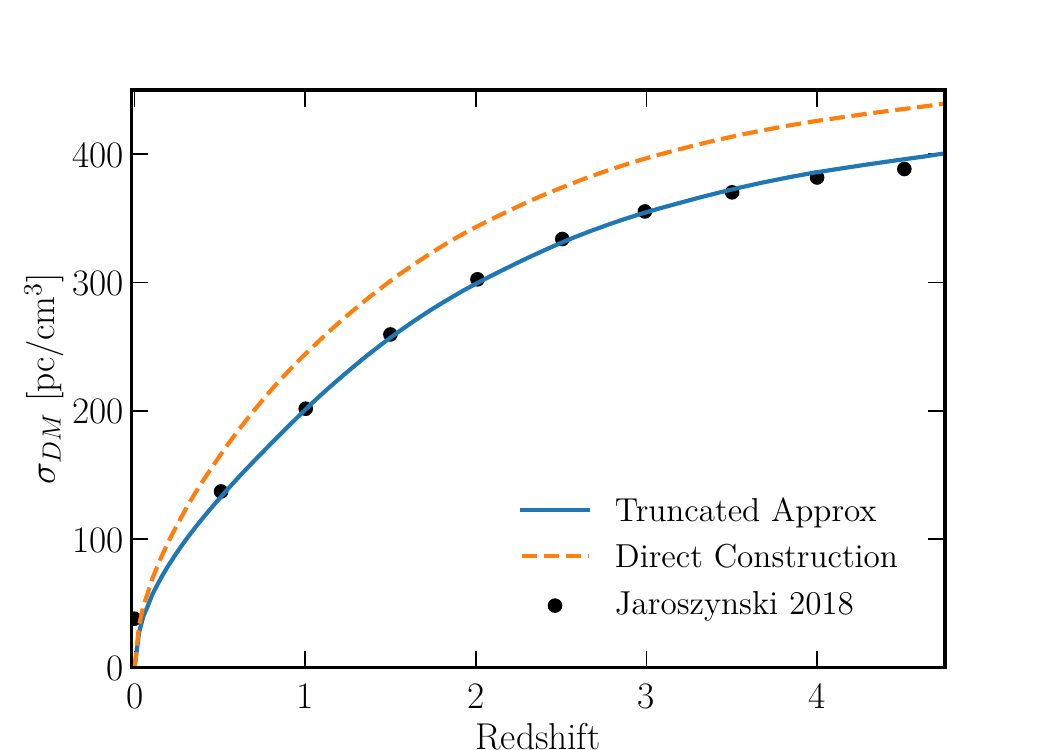}
    \caption{Mean and standard deviation of dispersion measures calculated from Illustris data. We compare the mean (left) and standard deviation (right) of the dispersion measures calculated in two ways using the Illustris data. The blue solid curve represents the results from our reproduction of the analysis technique used in \protect\citet{Jaroszynski_2019}, while the orange dashed curve represents the analysis technique used elsewhere in this paper, but applied to the Illustris data. For comparison, we plot the results from \protect\citet{Jaroszynski_2019} as black points. The mean dispersion measure of all three are consistent, while the standard deviation from \protect\citet{Jaroszynski_2019} and that calculated from our recreation of the same analysis are also consistent. However, there remains a $\sim 8$ per cent difference between the analysis technique used throughout this paper and the approach used in \protect\citet{Jaroszynski_2019}, compared to the $\sim 18$ per cent difference in Fig.~\ref{fig:DMstds}. This remaining difference can be attributed to the result of using different analyses, while the difference in Fig.~\ref{fig:DMstds} can be attributed to a combination of using different analyses and different datasets.}
    \label{fig:Illustris}
\end{figure*}

We calculate the total dispersion measure as detailed in the main body of the paper, and compare to the results from \citet{Jaroszynski_2019} in Fig.~\ref{fig:Illustris}. While both approaches predict similar mean dispersion measures, there is a notable difference between the standard deviations calculated using each approach. This difference can be traced entirely to the choice of range used in generating the histograms. Widening the range of the histograms can increase the standard deviation at any redshift, while reducing the range can substantially reduce the standard deviation.

We choose to use the approach that involves randomly selecting elements from the two snapshot distributions directly, without approximating them with histograms, as this seems to better support highly skewed distributions. While the contributions to the dispersion measure from the Illustris simulations can be approximated well by log-normal distributions, with only a slight skew, the same cannot be said of the CoDa~II--Dark Matter results, particularly at low redshifts. As seen in Fig.~\ref{fig:distcomp}, the contributions to the dispersion measure from the CoDa~II--Dark Matter simulation have a relatively large high-dispersion-measure tail, which may be related to the difference between the physics involved in the two simulations.

From this comparison of the analysis of Illustris data we can draw a few conclusions. First, and foremost, the majority of the difference between the standard deviations of the dispersion measure we predict from CoDa~II--Dark Matter and that arises in \citet{Jaroszynski_2019} does not come from the small difference in our analysis methods. Instead, it is a result of real differences in the Illustris and CoDa~II--Dark Matter data. Second, while the different approaches to interpolating between simulation snapshots is not the dominant source of difference in standard deviation, it does contribute. However, because of the skewness in the distribution of $(1+\delta_2)$ in the CoDa~II--Dark Matter results, the approach used in \citet{Jaroszynski_2019} cannot be directly applied to the CoDa~II--Dark Matter data.

\bsp
\label{lastpage}
\end{document}